# An empirical study to find an approximate ranking of citation statistics over subject fields

J. Martin van Zyl, Sean van der Merwe

*Department of Mathematical Statistics and Actuarial Science, University of the Free State, Bloemfontein, South Africa*

___________________________________________________________________________

**Abstract**

An empirical study is conducted to compare citations per publication, statistics and observed Hirsch indexes between subject fields using summary statistics of countries. No distributional assumptions are made and ratios are calculated. These ratios can be used to make approximate comparisons between researchers of different subject fields with respect to the Hirsch index.

___________________________________________________________________________

**1. Introduction**

Rating of researchers, and thus possibly also funding, is often based on the *h*-index or Hirsch index (Hirsch, 2005) which in turn is a function of the number of publications and citations received based on their publications. Impact numbers of journals are also based on citations. The purpose of this work is to show the differences between the average number of citations for various subject fields. The differences imply that researchers in certain fields will on average have higher *h*-index ratings than in other fields. Impact numbers of journals, irrespective of the period over which they are calculated, are a function of citations received for the papers published in the journal. Ideally, the various measures are meant to be used to compare within a subject field, but in practice the differences in, for example the number of citations between subject fields, are often not taken into account. A good overview with many references is given in the paper by Adler, Ewing and Taylor (2009). Much has been written on the use and misuse of impact numbers.

Scopus provides a data base where research output of countries in total and also per subject field for the period 1996 to 2010 is provided (SJR — SCImago Journal & Country Rank, 2007). This data was used in this work. It includes more than 10 000 000 citable documents and more than 100 000 000 citations and can be considered a very good approximation of all research results during that period, thus the population. Inference based on this data should be very close to the actual population parameters. There might have been changes in citation patterns during these years, but it is reasonable to assume that this bias will impact on all subject areas, and this will not influence the orderings. Scopus might be biased to include higher quality journals.

The ranking given in this work is an approximation, since it is based on the ratio of averages rather than the averages of ratios. In this research the assumption will be made that by using the ratios of two averages of citations per document of different subject fields, consistent orderings can be made, even though the approximation of the average of citations per individual document is only approximate. In other words, the estimated average number of citations per publication using totals is not a good approximation of the true average for individuals, but if ratios between subject fields are calculated using these estimates, then the ordering of the ratios is consistent and the ratios give a reasonable estimate of the true ratio calculated if the data for all individual researchers were available.

The summary statistics of countries and not of individual researchers are available. It can be shown that asymptotically these summary statistics can be used to estimate the average of individuals.

Part of the data available is $(p_j, c_j)$, $j = 1,...,n$, the number of citable documents and citations for a specific subject field and n=236 countries. Countries with research results in all subject areas were included. The number of citations per document and *h*-indexes for each subject field will be calculated, which can be used to make comparisons between individual researchers. A ratio of the average *h*-index between two subject fields is suggested as the measure to be used in a comparison. This applies on average and can be used as a guideline in such a comparison.

## 2. Summary statistics and confidence intervals

Ideally the average number of citations per publication over researchers should be used to make comparisons between subject fields, but only the totals of number of citations, total number of citable documents and the *h*-indexes of countries per subject and not individual results per researcher are available. The assumption is made that even though this is an approximation of citations per document for individual researchers, the ratios of these approximations with respect to the approximate average of citations per document and the *h*-indexes calculated over all subject fields will give a consistent ranking of subject fields relative to the average.

The asymptotic expected value of the ratio of two random variables is the ratio of the means, if the two variables both obey the weak law of large numbers. This principle supports the approximation of the average number of citations per author for a specific country, by using the ratio of the sum of citations to the number of publications of that specific country. For each country the data available is $(p_j, c_j)$, $j = 1,...,236$, the number of citable documents and citations for a specific subject field.

Consider a specific country, say country j, and a specific subject field with *m* researchers who each published $p_{ji}, i = 1,...,m$ citable publications and each researcher has $c_{ji}, i = 1,...,m$ citations associated with each publication. The sums $p_j = \sum_{i=1}^{m} p_{ji}$ and $c_j = \sum_{i=1}^{m} c_{ji}$ are available for $j = 1,...,n$ countries.

Using the results of Novak and Utev (1990) and the asymptotic distribution of ratios, it follows that for a specific country the following condition is fulfilled asymptotically:

$$\sum_{i=1}^{m} c_{ji} / \sum_{i=1}^{m} p_{ji} = m\bar{c}_j / m\bar{p}_j = \bar{c}_j / \bar{p}_j \approx E(c_j / p_j). \qquad (1)$$

Thus approximate expected values are calculated for country j, which are asymptotically equal to the true expected value of citations per document for researchers. It is assumed that even if the approximation of the average for an individual researcher is very weak, that the ordering of results will be consistent.

To calculate the average number of citations per document for a subject field the average is calculated over countries. The quality of research differs much between countries, thus to find an average over the whole spectrum of quality the plain sample mean and not a weighted mean was calculated to estimate results over countries for a specific subject field. Inference for a specific subject field over all countries was carried out using the estimated average citations per documents as calculated in (1).

Countries with at least one publication in the subject field were included to calculate the summary statistics. Confidence intervals for the ratios of specific subject fields with respect to the overall total research output will be calculated. The bootstrap confidence intervals are based on 10000 bootstraps. The results are given in table 1.

Citable documents and citations are heavy tailed distributed, but the ratio citations per document does not have very heavy tails and it is reasonable to assume that the mean of citations per document is finite and this variable obeys the weak law of large numbers. A histogram of citations per document for all countries with at least one citable document is shown in figure 1.

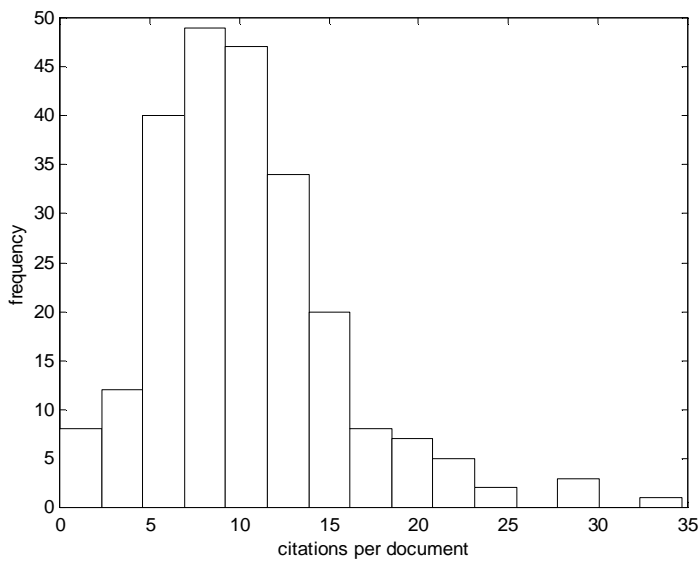

Figure 1. Histogram of number of citations per document over all subject areas of 236 countries.

In figure 2 it can be seen that the $h$-index is strongly dependent on the number of citations, which differs between subject fields, leading to higher $h$-factors in subject fields where one can expect more citations. This figure is made to illustrate the relationship using countries with complete data and zeros were not included, thus the line does not pass through the origin.

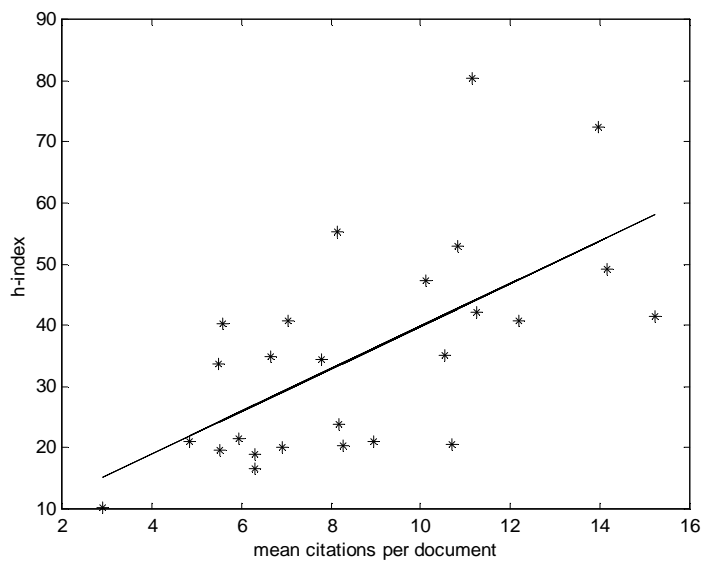

Figure 2. The relationship between the mean number of citations per subject field and the average *h*-index of countries.

The summary statistics of the approximated citations per document and average *h*-index per subject field are given in the appendix. These averages are used to calculate ratios with respect to averages over all subject fields. There are big differences between subject fields with respect to the citations per document and the *h*-indexes.

In table 1 ratios with confidence intervals for the mean number of citations per document divided by the mean number of citations per document over all subject fields are given for a selection of subject fields. The full table is given in the appendix. The mean for all subject fields was calculated by using as sample the results of the data of countries over all subject areas. A ratio of one will thus be in line with the average, while a ratio of more than one indicates a subject area which receives more citations per document than the average.

The multidisciplinary subject field is a total outlier compared to the other results with respect to average number of citations per published paper. Medically related subject fields have higher ratios and Medicine, Biochemistry, Genetics, Molecular Biology, Immunology, Microbiology and Neuroscience are above the average. For the citations per document Neuroscience, with a ratio of 1.59 relative to the average over all subject areas, and for the *h*-index Medicine, with a ratio of 2.66, yielded the highest ratios relative to the average.

|  | Citations per document (cpd) | | | h-index | | |
| --- | --- | --- | --- | --- | --- | --- |
| Subject Area | 2.5% quantile | mean cpd ratio vs all | 97.5% quantile | 2.5% quantile | Mean h ratio vs all | 97.5% quantile |
| All |  | 1.0 |  |  | 1.0 |  |
| Agricultural and Biological Sciences | 0.91 | 0.94 | 0.98 | 1.73 | 1.85 | 1.98 |
| Biochemistry, Genetics and Molecular Biology | 1.40 | 1.45 | 1.51 | 2.02 | 2.08 | 2.14 |
| Chemistry | 1.03 | 1.09 | 1.14 | 1.39 | 1.46 | 1.53 |
| Economics, Econometrics and Finance | 0.52 | 0.57 | 0.62 | 0.43 | 0.46 | 0.49 |
| Immunology and Microbiology | 1.52 | 1.59 | 1.66 | 1.61 | 1.73 | 1.87 |
| Mathematics | 0.48 | 0.54 | 0.60 | 0.79 | 0.86 | 0.92 |
| Medicine | 1.10 | 1.16 | 1.21 | 2.54 | 2.66 | 2.79 |
| Physics and Astronomy | 0.80 | 0.87 | 0.96 | 1.25 | 1.37 | 1.51 |
| Psychology | 0.77 | 0.85 | 0.93 | 0.46 | 0.50 | 0.54 |

Table 1. Selection of mean ratios of citations per publication in a subject area to citations in all subject areas, and mean ratio of *h*-index per subject area to *h*-index over all areas. The lower and upper limits of a 95% bootstrap confidence interval are given with the mean ratio.

These ratios can be used in the following way: Say a researcher in Agricultural and Biological Sciences has a *h*-index of 20, and a researcher in Mathematics a *h*-index of 15, to be comparable on the mathematics level, the h-factor of 20 should be multiplied by 0.86/1.85 giving a result of 9.2930, which shows that when taking the differences in expected citations between subjects into account, the mathematician is performing better with respect to the *h*-index. The ranking of the ratios of individual subject fields for citations per document and the *h*-index to the all subject fields combined are shown in figure 5 and figure 6.

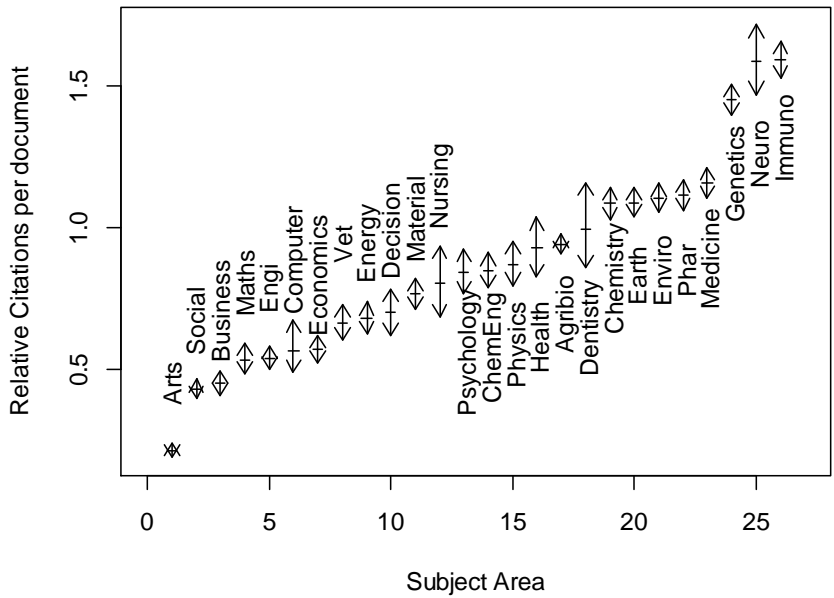

Figure 3. Mean ratio with 95% confidence interval of citations per documents per subject area to citations per document over all subject areas.

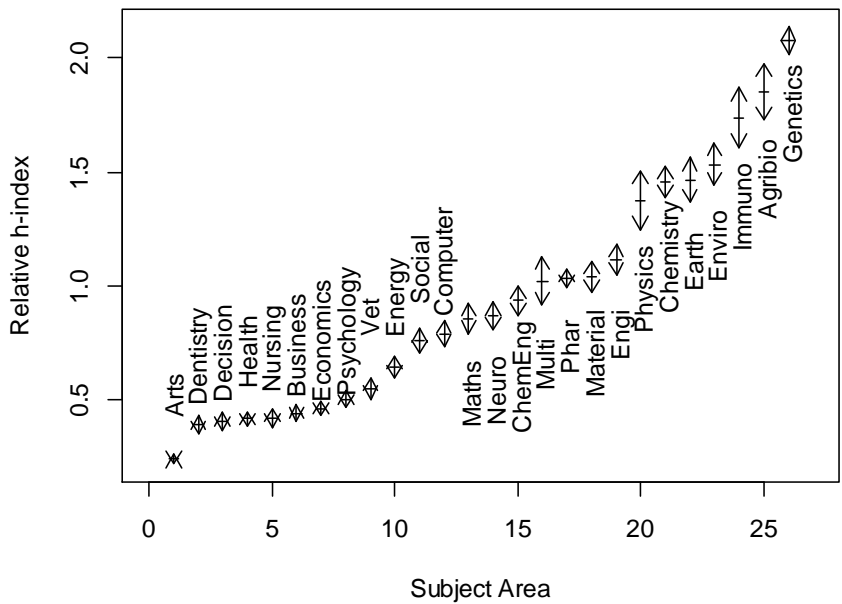

Figure 4. Mean ratio with 95% confidence interval of $h$-index per subject area to h-index of all subject areas. Medicine with a ratio of 2.66 is not shown in the figure.

In order to confirm that different groupings are formed with respect to the different subject groups, the multivariate technique dimensional scaling, using a metric solution and the correlation matrix was performed. The variables are the different subject fields and the observations the vectors per subject field of respectively citations per document and for the second analysis the matrix of $h$-indexes of countries.

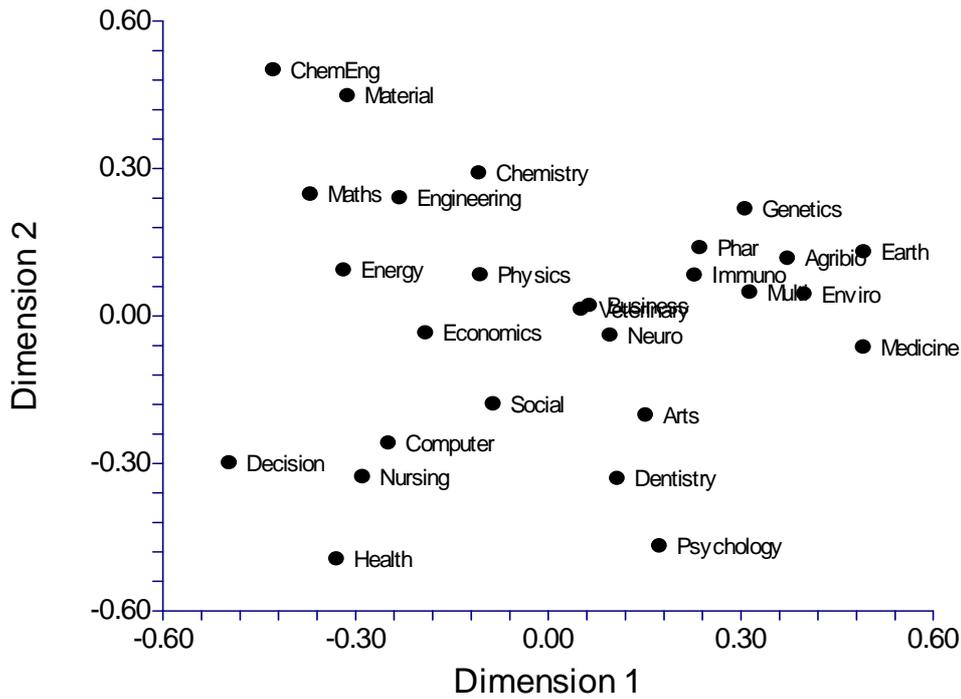

Figure 5. Multidimensional scaling plot showing dimensions 1 and 2, using citations per document for various subject fields.

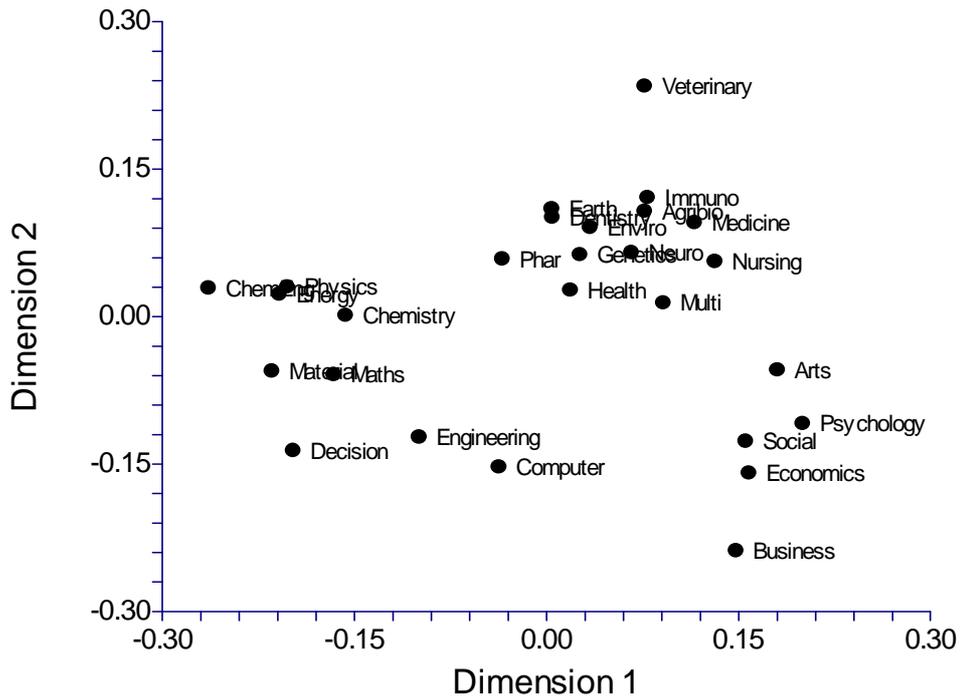

Figure 6. Multidimensional scaling plot showing dimensions 1 and 2, using the correlation matrix between *h*-indexes for various subject fields.

It is clear in both figures that groupings are formed and it is not a homogeneous group with respect to citations per document and the Hirsch index ratings. Roughly three groups can be observed, life sciences, the second group related to behavioural patterns of humans and then physical sciences. This is in line with the ordering found using ratios where the medical and biological sciences have the highest ratios with respect to citation statistics, other physical sciences more or less in the middle and behavioural subjects have on average the smallest ratios. In other words the groups are also formed approximately if one looks at the ordering of the ratios.

## 3. Conclusions

The calculated ratios are averages and can be used as a guideline to make comparisons between the research output of different subject fields. Thus it is not exact and only an approximation. It may also be that the ratios differ if only top researchers are considered as opposed to including the whole sample.

Ultimately, no matter how the calculations are performed, it is clear that there exists a large difference in the way research is cited between subject fields. The differences are too large to be explained by only the fact that some fields are more 'relevant' or 'interesting' than others. It seems that part of the differences must be accounted for by distinct citation cultures. This argument is supported by the separation evident in figure 6.

Any attempt to rate researchers or journals using a single measure or unified benchmark system across subject fields is thus inherently biased towards fields with a natural culture of high numbers of citations per document. Benchmarks should be limited to one field, or an attempt must be made to adjust for the field of research when rating researchers and journals.

Even though there are weaknesses in this approximation, it can be invaluable as a guideline to make reasonable comparisons when evaluating researchers, departments, institutions and also journals across subject fields.

## Appendix

|   | Subject Area | 2.5% quantile of ratio (citations) | cpd ratio vs all (citations) | 97.5% quantile of ratio (citations) | 2.5% quantile of ratio | $h$-index ratio vs all | 97.5% quantile of ratio |
|---|---|---|---|---|---|---|---|
|   | All Subjects |   | 1.0 |   |   | 1.0 |   |
| 1 | Agricultural and Biological Sciences | 0.91 | 0.94 | 0.98 | 1.73 | 1.85 | 1.98 |
| 2 | Arts and Humanities | 0.19 | 0.22 | 0.24 | 0.22 | 0.24 | 0.26 |
| 3 | Biochemistry, Genetics and Molecular Biology | 1.40 | 1.45 | 1.51 | 2.02 | 2.08 | 2.14 |
| 4 | Business, Management and Accounting | 0.41 | 0.45 | 0.50 | 0.41 | 0.44 | 0.48 |

| # | Field | | | | | |
|---|---|---|---|---|---|---|
| 5 | Chemical Engineering | 0.79 | 0.85 | 0.91 | 0.87 | 0.93 | 1.00 |
| 6 | Chemistry | 1.03 | 1.09 | 1.14 | 1.39 | 1.46 | 1.53 |
| 7 | Computer Science | 0.49 | 0.57 | 0.67 | 0.73 | 0.79 | 0.85 |
| 8 | Decision Sciences | 0.62 | 0.70 | 0.78 | 0.37 | 0.40 | 0.44 |
| 9 | Dentistry | 0.86 | 0.99 | 1.16 | 0.35 | 0.39 | 0.43 |
| 10 | Earth and Planetary Sciences | 1.04 | 1.09 | 1.14 | 1.37 | 1.46 | 1.56 |
| 11 | Economics, Econometrics and Finance | 0.52 | 0.57 | 0.62 | 0.43 | 0.46 | 0.49 |
| 12 | Energy | 0.63 | 0.68 | 0.74 | 0.59 | 0.64 | 0.69 |
| 13 | Engineering | 0.50 | 0.54 | 0.58 | 1.05 | 1.11 | 1.18 |
| 14 | Environmental Science | 1.06 | 1.10 | 1.16 | 1.44 | 1.53 | 1.63 |
| 15 | Health Professions | 0.83 | 0.93 | 1.04 | 0.38 | 0.41 | 0.45 |
| 16 | Immunology and Microbiology | 1.52 | 1.59 | 1.66 | 1.61 | 1.73 | 1.87 |
| 17 | Materials Science | 0.72 | 0.77 | 0.83 | 0.97 | 1.04 | 1.11 |
| 18 | Mathematics | 0.48 | 0.54 | 0.60 | 0.79 | 0.86 | 0.92 |
| 19 | Medicine | 1.10 | 1.16 | 1.21 | 2.54 | 2.66 | 2.79 |
| 20 | Multidisciplinary | 3.43 | 3.85 | 4.28 | 0.92 | 1.02 | 1.13 |
| 21 | Neuroscience | 1.47 | 1.59 | 1.72 | 0.80 | 0.87 | 0.93 |
| 22 | Nursing | 0.69 | 0.81 | 0.94 | 0.38 | 0.42 | 0.46 |
| 23 | Pharmacology, Toxicology and Pharmaceutics | 1.06 | 1.11 | 1.17 | 0.99 | 1.03 | 1.07 |
| 24 | Physics and Astronomy | 0.80 | 0.87 | 0.96 | 1.25 | 1.37 | 1.51 |
| 25 | Psychology | 0.77 | 0.85 | 0.93 | 0.46 | 0.50 | 0.54 |
| 26 | Social Sciences | 0.40 | 0.43 | 0.47 | 0.70 | 0.76 | 0.81 |
| 27 | Veterinary | 0.61 | 0.67 | 0.73 | 0.50 | 0.55 | 0.60 |